\documentclass[twocolumn,showpacs,preprintnumbers,amsmath,amssymb]{revtex4}

\usepackage{graphicx}
\usepackage{dcolumn}
\usepackage{amsmath}
\usepackage{amssymb}

\newcommand{\bvec}[1]{\mbox{\boldmath $#1$}}
\newcommand{\D}{\delta }

\def\vq{{\boldsymbol q}}

\def\vk{{\boldsymbol k}}

\def\vS{{\boldsymbol S}}

\newcommand{\eq}[1]{Eq.~(\ref{#1})}
\newcommand{\fig}[1]{Fig.~\ref{#1}}
\newcommand{\be}{\begin{equation}}
\newcommand{\ee}{\end{equation}}
\newcommand{\bea}{\begin{eqnarray}}

\newcommand{\eea}{\end{eqnarray}}
\newcommand{\bean}{\begin{eqnarray*}}
\newcommand{\eean}{\end{eqnarray*}}
\newcommand{\bfi}{\begin{figure}}
\newcommand{\efi}{\end{figure}}
\newcommand{\bc}{\begin{center}}
\newcommand{\ec}{\end{center}}
\newcommand{\ba}{\begin{array}}
\newcommand{\ea}{\end{array}}

\begin{document}


\title{Excitation spectrum of {$\boldsymbol d$}-wave Fermi surface deformation}

\author{Hiroyuki Yamase} 
\affiliation{RIKEN (The Institute of
Physical and Chemical Research), Wako, Saitama 351-0198, Japan \\
Max-Planck-Institute for Solid State Research, Heisenbergstrasse 1, 
D-70569 Stuttgart, Germany}


\date{\today}

\begin{abstract}
Several instabilities competing with  the $d$-wave singlet pairing were 
proposed for high-$T_{\rm c}$ cuprates. 
One of them is the $d$-wave Fermi surface deformation ($d$FSD), 
which is generated by forward scattering. 
In this paper, 
correlation functions of the $d$FSD 
are calculated within the random phase approximation.   
In the normal state, the excitation spectrum shows a low energy peak, 
which smoothly connects to critical fluctuations of the $d$FSD at 
lower temperature. 
The competition with the $d$-wave pairing, however, blocks 
the critical fluctuations. 
The whole spectral weight is transferred to high energy and 
a pronounced peak appears there in the $d$-wave pairing state. 
This peak is an overdamped collective mode of the $d$FSD 
and can grow to be  a 
resonance mode at moderate finite wavevectors.  
\end{abstract}

\pacs{71.18.+y, 71.10.Fd, 74.25.-q, 74.72.-h}
\maketitle

High-$T_{\rm c}$ cuprates are doped Mott insulators. 
The parent compounds are antiferromagnetic Mott insulators, 
which become high-$T_{\rm c}$ superconductors with carrier doping. 
The superconducting state does not have an isotropic gap such as 
BCS superconductors, but has an anisotropic gap with the  
$d$-wave symmetry.

It has been recognized that there 
may be several instabilities competing with the 
$d$-wave superconductivity. 
Effects of the antiferromagnetism may play a crucial role still in 
the $d$-wave superconducting state, and their competition was  
discussed using a concept of the SO(5) symmetry.\cite{zhang97,demler04} 
Another idea, a self-organized one-dimensional 
charge order in the CuO$_{2}$ plane --- spin-charge stripes 
hypothesis, was proposed to discuss several 
experimental data\cite{tranquada95,kivelson03} 
and phenomenological theories 
were developed.\cite{emery97} 
Through a microscopic analysis of the two-dimensional (2D) $t$-$J$ model 
by $1/N$ expansions for Hubbard operators, 
the $d$-wave charge density order was proposed.\cite{zeyher99}   
This phase has bond currents forming staggered flux, and was 
discussed in different contexts.\cite{varma97,chakravarty01}
A similar state, called the staggered flux phase, was 
proposed in 
the SU(2) slave-boson formalism in the $t$-$J$ model.\cite{wen96}  
In this scheme, the exact SU(2) gauge symmetry 
at half-filling\cite{affleck88} was invoked also at finite doping; 
the underlying theoretical concept is quite different 
from the $d$-wave charge density order.  
All these possible competing orders come from 
electron-electron correlations with large momentum transfer 
near $\vq=(\pi,\,\pi)$.  

Recently another competing order was 
proposed,\cite{yamase00,metzner00} 
the $d$-wave Fermi surface deformation ($d$FSD).\cite{miscdFSD}  
The Fermi surface (FS) expands along the $k_{x}$-direction 
and shrinks along the $k_{y}$-direction (or vice versa). 
This order has to be distinguished from the above possible 
competing orders.  
The channel of this instability is forward scattering 
with $\vq=(0,\,0)$. The $d$FSD was first discussed for 
the 2D $t$-$J$ model\cite{yamase00} and Hubbard model.\cite{metzner00} 
It was tested in several renormalization group schemes 
applied to the Hubbard model.\cite{metzner00,wegner02,honerkamp02} 
The $d$FSD was investigated also in perturbation theories for 
the Hubbard model\cite{frigeri02,neumayr03},   
in the mean-field theory for the 
extended Hubbard model,\cite{valenzuela01} and 
in phenomenological models.\cite{metzner03,khavkine04,yamase04b} 
In the continuum (not lattice) model FS deformation was investigated 
in analogy to the nematic phase 
in liquid crystals.\cite{oganesyan01,kee03} 

In accord with a result in the Hubbard model,\cite{honerkamp02}  
the analysis of the $t$-$J$ model\cite{yamase00} showed that 
an instability of the $d$FSD competed with  
a more dominant instability, the $d$-wave singlet pairing, 
and was usually masked and not seen. 
However, it was shown that the presence of a small extrinsic anisotropy 
was sufficient to manifest the $d$FSD.\cite{yamase00}  
This implies that while the spontaneous instability of the $d$FSD does 
not take place, the electron system still has an appreciable susceptibility 
of the $d$FSD and is sensitive to the external anisotropy; 
the FS is softened.\cite{metzner03} 
This idea was invoked for LSCO systems through the consideration of 
band parameter dependences\cite{yamase00} 
and magnetic excitation spectra.\cite{yamase010203}

In this letter, we investigate dynamical properties of the $d$FSD. 
Since the instability of the $d$FSD is signaled by divergence of 
its static susceptibility at $\vq=0$, 
we focus on the dynamical susceptibility near $\vq=0$ 
and calculate it within the random phase approximation (RPA). 
In the normal state the excitation spectrum shows a 
low energy peak, which smoothly connects to critical fluctuations of the 
$d$FSD at lower temperature. The critical fluctuations are, however, 
blocked by the more dominant $d$-wave pairing instability. 
The low energy spectral weight 
is suppressed and vanishes at zero temperature. Instead the spectral weight 
is transfered to high energy and we find a pronounced peak there. 
This peak is an overdamped collective mode of the $d$FSD 
and can grow to be  
a resonance mode at moderate finite wavevectors.

To investigate correlations of the $d$FSD, we take the 2D $t$-$J$ model on 
the square lattice, 
\be
 H = -  \sum_{i,\,j,\, \sigma} t^{(l)}
 \tilde{c}_{i\,\sigma}^{\dagger}\tilde{c}_{j\,\sigma}+
   J \sum_{\langle i,j \rangle} \vS_{i} \cdot \vS_{j},  \label{tJ} 
\ee  
defined in the Fock space with no doubly occupied sites. 
Here $\tilde{c}_{i\,\sigma}$ ($\vS_{i}$) is an electron (a spin) 
operator.  
The $t^{(l)}$ is the $l$th ($l \leq 2$) 
neighbor hopping integral, and we denote $t^{(1)}=t$, $t^{(2)}=t'$. 
The $J(>0)$ is the superexchange coupling between nearest-neighbor sites. 
We introduce U(1) slave-particles as 
$\tilde{c}^{\dagger}_{i\,\sigma}=f_{i\,\sigma}^{\dagger}b_{i}$,  
where $f_{i\,\sigma}$ ($b_{i}$) is a fermion (boson) operator  
that carries spin $\sigma$ (charge $e$),  and 
$\vS_{i} = \frac{1}{2} f_{i\, \alpha}^{\dagger}\bvec{\sigma}_{\alpha\,\beta} 
f_{i\, \beta}$ with Pauli matrix $\bvec{\sigma}$. This is 
an exact transformation. 
We then decouple the interactions with the so-called 
resonating-valence-bond mean fields:  
$\chi_{\tau}$$\equiv$$\langle \sum_{\sigma}f_{i\,\sigma}^{\dagger}
f_{i+\tau \,\sigma}\rangle$, 
$\langle b_{i}^{\dagger}b_{i+\tau}\rangle$, and 
$\Delta_{\tau}$$\equiv$$\langle f_{i\,\uparrow}f_{i+\tau \,\downarrow}- 
f_{i\,\downarrow}f_{i+\tau \,\uparrow}\rangle$, where 
$\boldsymbol\tau=\bvec{r}_{j}-\bvec{r}_{i}$ denotes the direction. 
These mean fields are assumed to be  
real constants independent of sites $i$. 
We approximate the boson to condense  
at the bottom of its band, which is reasonable at low temperature $T$, 
and obtain the following Hamiltonian:  
\be
\hspace{-0mm}H_{0}=\sum_{\vk}
\left(
      f_{\vk\,\uparrow}^{\dagger}\;\; f_{-\vk\,\downarrow}
\right)
\left( \begin{array}{cc} 
   \xi_{\vk} & -\Delta_{\vk} \\
-\Delta_{\vk} & -\xi_{\vk}
          \end{array}\right)
\left( \begin{array}{c}
 f_{\vk\,\uparrow} \\
 f_{-\vk\,\downarrow}^{\dagger}
\end{array}\right)\, ,\label{MFH}
\ee
with a global constraint $\sum_{\sigma}\langle 
f^{\dagger}_{i\sigma}f_{i\sigma}\rangle =1-\D$. 
Here $\xi_{\vk}=-2
\left(F_{x}\cos k_{x}+F_{y}\cos k_{y}
+ 2t'\D \cos k_{x} \cos k_{y}\right) 
-\mu$, $\Delta_{\vk}=-\frac{3}{4}J 
 \left(\Delta_{x}\cos k_{x}+\Delta_{y} \cos k_{y}\right)$, and  
$F_{x(y)}=t\D+\frac{3}{8}J \chi_{x(y)}$, 
with $\D$ being hole density and $\mu$ the chemical potential. 
The mean fields are determined self-consistently by minimizing 
the free energy. 
The isotropic state $\chi_{x}=\chi_{y}$ is stabilized and the 
$d$-wave singlet pairing, $\Delta_{x}=-\Delta_{y}=\Delta_{0}$,  
sets in at low $T$.

The advantages of this formalism\cite{fukuyama98} are 
(i) the phase diagram on $T$ versus $\D$ 
catches essential physics of high-$T_{\rm c}$ cuprates, 
(ii) magnetic excitation in 
actual systems was consistently described  on the basis of fermiology 
including material 
dependence,\cite{tanamoto94,brinckmann01,yamase010203} and 
(iii) the $d$FSD channel was shown to exist in 
the $J$-term,\cite{yamase00} which enable us study its 
competition with the $d$-wave pairing on an 
equal footing.

To analyze correlations of the $d$FSD, we define 
$d$-wave weighted fermion density,  
\be
\hat{\chi}_{d}(\vq) =\sum_{\vk\,\sigma}d_{\vk}
f_{\vk-\frac{\vq}{2}\, \sigma}^{\dagger}
f_{\vk+\frac{\vq}{2}\, \sigma}\,,\label{chid} 
\ee 
with $d_{\vk}=\frac{1}{2}(\cos k_{x}- \cos k_{y})$. 
The spontaneous $d$FSD is described by 
$\langle \hat{\chi}_{d}({\boldsymbol 0})\rangle \neq 0$, which is, however, 
prohibited by competition with the 
$d$-wave singlet pairing. 
In the $d$-wave singlet state, fluctuations of $\hat{\chi}_{d}(\vq)$ 
induce fluctuations of extended $s$-wave pairing.\cite{yamase00}  
To include this effect we also define 
\be
\hspace{-0mm}\hat{\Delta}_{s}(\vq)=\sum_{\vk}s_{\vk}
\left(f_{\vk+\frac{\vq}{2}\, \uparrow}f_{-\vk+\frac{\vq}{2}\, \downarrow} 
-f^{\dagger}_{\vk-\frac{\vq}{2}\, \uparrow}
f^{\dagger}_{-\vk-\frac{\vq}{2}\, \downarrow}\right)\, ,\label{deltas} 
\ee
with $s_{\vk}=\frac{1}{2}(\cos k_{x}+\cos k_{y})$. 
Thus the correlation function forms a $2\times 2$ matrix, 
\bea
{\boldsymbol{\kappa}_{0}}(\vq,\,\omega) = \left(
\begin{array}{cc}
\kappa_{0}^{11} (\vq\,,\omega) & \kappa_{0}^{12} (\vq\,,\omega) \\
\kappa_{0}^{21} (\vq\,,\omega) & \kappa_{0}^{22} (\vq\,,\omega) 
\end{array}
\right)\,,
\eea
where $\kappa_{0}^{12} (\vq\,,\omega)=\kappa_{0}^{21} (\vq\,,\omega)$, and 
\bea
&&\hspace{-10mm}\kappa_{0}^{11}(\vq,\,\omega) 
= \frac{{\rm i}}{N}\int_{0}^{\infty} {\rm d}t 
{\rm e}^{{\rm i}(\omega+{\rm i}\Gamma) t}
\langle 
[\hat{\chi}_{d}(\vq,\,t), \hat{\chi}_{d}(-\vq)]\rangle_{0}\,, 
\label{k11}\\
&&\hspace{-10mm} \kappa_{0}^{12}(\vq,\,\omega) 
= \frac{{\rm i}}{N}\int_{0}^{\infty} {\rm d}t   
{\rm e}^{{\rm i}(\omega+{\rm i}\Gamma) t}
\langle 
[\hat{\chi}_{d}(\vq,\,t), \hat{\Delta}_{s}(-\vq)]\rangle_{0}\,,
\label{k12} \\
&&\hspace{-10mm} \kappa_{0}^{22}(\vq,\,\omega) 
= \frac{{\rm i}}{N}\int_{0}^{\infty} {\rm d}t 
{\rm e}^{{\rm i}(\omega+{\rm i}\Gamma) t} 
\langle 
[\hat{\Delta}_{s}(\vq,\,t), \hat{\Delta}_{s}(-\vq)]\rangle_{0}
\label{k22}\,. 
\eea
Here $\hat{\chi}_{d}(\vq,\,t)={\rm e}^{{\rm i}H_{0}t} 
\hat{\chi}_{d}(\vq) {\rm e}^{{\rm -i}H_{0}t}$, and 
$\hat{\Delta}_{s}(\vq,\,t)={\rm e}^{{\rm i}H_{0}t} 
\hat{\Delta}_{s}(\vq) {\rm e}^{{\rm -i}H_{0}t}$. 
The bracket $\langle \cdots \rangle_{0}$ denotes 
an expectation value under the 
Hamiltonian (\ref{MFH}),  
and $[\cdot\, , \,\cdot]$ is the commutator; 
$N$ is the total number of lattice sites. 
We consider interactions in the RPA, 
\be
\left(
\begin{array}{cc}
\kappa^{11} & \kappa^{12} \\
\kappa^{21} & \kappa^{22} 
\end{array}
\right)^{-1}=
\left(
\begin{array}{cc}
\kappa^{11}_{0} & \kappa^{12}_{0} \\
\kappa^{21}_{0} & \kappa^{22}_{0}
\end{array}
\right)^{-1}-
\left(
\begin{array}{cc}
3J/2 &0 \\
0& 3J/2 
\end{array}
\right)\,. \label{RPA}
\ee

\begin{figure}[t]
%
\centerline{\includegraphics[scale=0.65]{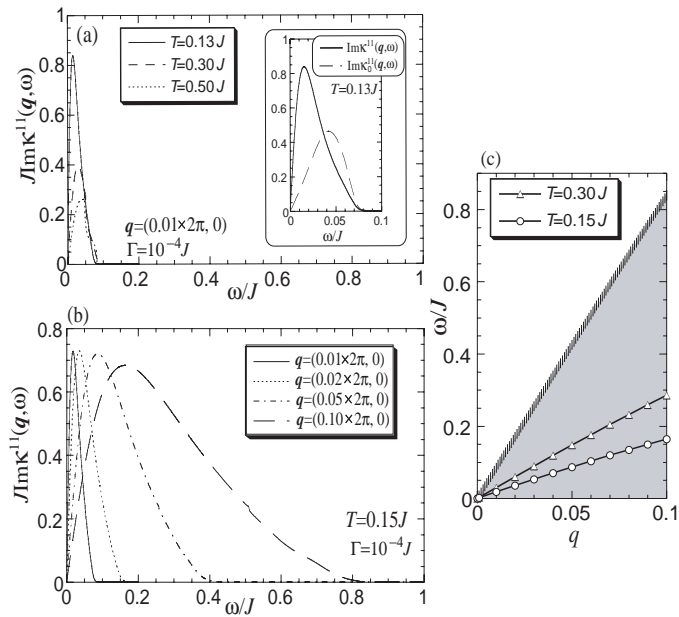}}
%
\caption{Normal state. 
(a) $\omega$ dependence of Im$\kappa^{11}(\vq,\,\omega)$ 
for several choices of $T$ at $\vq=(0.01\times 2\pi,\,0)$. 
The result at $T=0.13J$ is compared with 
Im$\kappa^{11}_{0}(\vq,\,\omega)$ in the inset.
(b) $\omega$ dependence of Im$\kappa^{11}(\vq,\,\omega)$ 
for several choices of $\vq$ along the $[10]$ direction. 
(c) Excitation spectrum on the plane of $\omega$ vs. $q$. 
The shaded region is a particle-hole continuum. 
The open circle (triangle) corresponds to 
the peak energy of Im$\kappa^{11}(\vq,\,\omega)$ 
at a given $q$ at $T=0.15J$ $(0.30J)$; 
$q$ is defined as $\vq=(q \times 2\pi,\,0)$. 
}
\label{uImkqw}
\end{figure}

\begin{figure}[t]
%
\centerline{\includegraphics[scale=0.78]{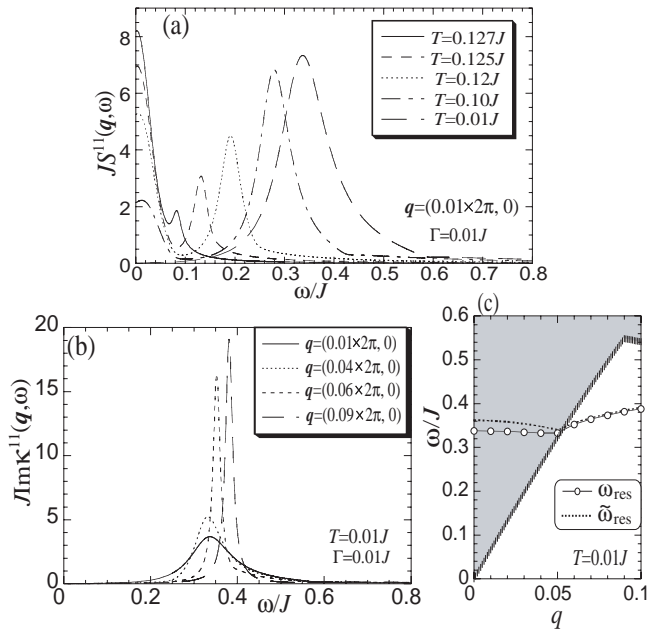}}
%
\caption{$d$-wave pairing state. 
(a) $\omega$ dependence of $S^{11}(\vq,\,\omega)$ 
for several values of $T$ at $\vq=(0.01 \times 2\pi,\,0)$. 
(b) $\omega$ dependence of Im$\kappa^{11}(\vq,\,\omega)$ 
for several choices of $\vq$ along the $[10]$ direction ($T=0.01J$). 
(c) Excitation spectrum on the plane of $\omega$ vs. $q$. 
The shaded region is a continuum of excitations.
The open circle corresponds to the peak energy of 
Im$\kappa^{11}(\vq,\,\omega)$, 
$\omega_{\rm res}$, at a given $q$ ($T=0.01J$); 
$q$ is defined as $\vq=(q \times 2\pi,\,0)$. 
$\tilde{\omega}_{\rm res}$ is estimation by \eq{resonance}. 
}
\label{dImkqw}
\end{figure}

In this letter, we focus on 
$\kappa^{11}(\vq,\,\omega)$ with $\vq \approx 0$, and 
investigate its spectral weight in both 
the normal state and the $d$-wave singlet pairing state. 
Full results including other components of $\boldsymbol{\kappa}$ will be 
shown elsewhere.\cite{yamase04c}  
We take band parameters, $t/J=4$ and $t'/t=-1/6$, for which  
the $d$FSD is known to be prominent;\cite{yamase00} 
a doping rate is fixed to $\D=0.10$. 
In Eqs.(\ref{k11})-(\ref{k22}), the value of $\Gamma$ is a positive 
infinitesimal and we take $\Gamma=10^{-4}J$ (\fig{uImkqw}) or 
$\Gamma=0.01J$ (\fig{dImkqw}) in numerical calculations.


Figure~\ref{uImkqw}(a) shows Im$\kappa^{11}(\vq,\,\omega)$ as 
a function of $\omega$ for several choices of $T$ 
in the normal state at $\vq=(0.01\times 2\pi,\,0)$. 
The spectral weight concentrates at low energy for all $T$. 
This is due to a property of Im$\kappa^{11}_{0}(\vq,\,\omega)$ 
that particle-hole excitations obey the relation, 
$\omega=\xi_{\vk+\vq/2}-\xi_{\vk-\vq/2}$; 
$\omega$ becomes small for small $\vq$. 
The low energy spectral weight increases with decreasing $T$ and the 
peak position shifts closer to zero energy. 
This enhancement comes from the interactions in the RPA, \eq{RPA}, 
and smoothly connects to critical fluctuations of the $d$FSD at lower $T$. 
While the present lowest temperature ($T=0.13J$) is much 
higher than the critical temperature of the $d$FSD 
($T_{d\text{FSD}}=0.038J$), the low energy weight of 
Im$\kappa^{11}(\vq,\,\omega)$ substantially increases in 
comparison with Im$\kappa_{0}^{11}(\vq,\omega)$ as shown in 
the inset of \fig{uImkqw}(a). 
In this sense, the enhancement 
of the low energy peak is a precursor of collective 
fluctuations of the $d$FSD. 
In \fig{uImkqw}(b), we plot Im$\kappa^{11}(\vq,\,\omega)$ 
for several choices of  $\vq\ (\parallel [10])$ at given $T$. 
The spectral weight spreads to 
higher energy with increasing $|\vq|$, 
but the peak position stays at relatively low energy, which is  
due to the enhancement by the RPA. 
In \fig{uImkqw}(c), we summarize the excitation spectrum on the plane of 
$\omega$ vs. $q$. The shaded region is a gapless particle-hole 
continuum and the upper edge increases linearly with  $q$.  
The peak energy of Im$\kappa^{11}(\vq,\,\omega)$ 
disperses linearly with $q$ at low $q$ within numerical accuracy. 
The gradient of the $q$-linear becomes small at low $T$, which is 
due to the enhancement of low energy fluctuations of the $d$FSD. 
These qualitative features have been checked also along 
$\vq \parallel [11]$.

Further decreasing $T$ below values shown in 
\fig{uImkqw}(a), the $d$-wave singlet pairing instability takes place, 
which competes with the $d$FSD and prohibits the spontaneous $d$FSD. 
This competition is shown in \fig{dImkqw}(a); 
we take the spectral function, 
$S^{11}(\vq,\,\omega) = 2 {\rm Im}\kappa^{11}(\vq,\,\omega) 
/ (1- {\rm e}^{-\omega/ T})$, to see low energy structures also 
on the same scale,   
and plot its $\omega$ dependence for 
several choices of $T$. 
The low energy weight 
is suppressed with decreasing $T$  
and is transferred to higher energy 
to form a second peak there (see the results for $T \gtrsim 0.12J$). 
While the low energy weight vanishes at $T=0$, 
the second peak grows to be a pronounced peak at low $T$. 
To see its dispersive features, 
we calculate the $\omega$ dependence of Im$\kappa^{11}(\vq,\,\omega)$ 
for several choices of $\vq$ at low $T$. 
Figure~\ref{dImkqw}(b) shows that the peak width becomes narrower 
with $|\vq|$ and a sharp peak appears at moderate 
$|\vq|$ ($\gtrsim 0.05 \times 2\pi$). 
This is a resonance peak and an in-gap collective mode 
of the $d$FSD, namely a bound state  
(the finite peak width of the resonance is due to $\Gamma >0$ 
in the numerical calculations).  
To see this, we calculate the gap energy 
of Im$\kappa^{11}_{0}(\vq,\,\omega)$, namely a lower edge of a 
continuum of excitations, and plot it as a function of $q$ 
as well as the peak energy of Im$\kappa^{11}(\vq,\,\omega)$, 
$\omega_{\rm res}$, in \fig{dImkqw}(c). 
The lower edge increases with $q$  
and $\omega_{\rm res}$ is located inside the gap (in $q \gtrsim 0.05$), 
which gives rise to the resonance peak. 
This resonance does not appear along $\vq \parallel [11]$ at least up to 
$\vq =(0.10 \times 2\pi, \, 0.10 \times 2\pi)$ because of the 
extention of the continuum spectrum down to lower energy.

To understand the dispersion relation of the resonance, 
we first approximate $\kappa^{11}(\vq,\,\omega)$ given by 
\eq{RPA} to $\kappa^{11}(\vq,\,\omega)=\kappa_{0}^{11} (\vq,\,\omega) 
/(1-3J\kappa_{0}^{11}(\vq,\,\omega)/2)$; 
we have checked numerically that quantitative changes 
by this approximation are not noticeable. 
The dispersion is then determined 
by the condition, 
\be 
1-\frac{3}{2}J{\rm Re}\kappa_{0}^{11}(\vq,\,\tilde{\omega}_{\rm res})=0\,.
\label{resonance}
\ee 
This equation can have two solutions for a given $\vq$ and the 
$\tilde{\omega}_{\rm res}$ is the smaller one. 
We solve \eq{resonance} numerically and compare 
$\tilde{\omega}_{\rm res}$ with $\omega_{\rm res}$ in \fig{dImkqw}(c). 
We see a good agreement in a moderate $q$-region, 
where the resonance appears. 
A poor agreement in a small $q$-region is due to finite 
weight of Im$\kappa_{0}^{11}(\vq,\,\omega)$, 
which invalidates using \eq{resonance}. 
We, however, see that the dispersive features are well 
characterized by \eq{resonance} in a whole $q$-region in 
\fig{dImkqw}(c).

It should be noted that \eq{resonance} has solutions for any 
$\vq$ shown in \fig{dImkqw}(c). 
This is due to the enhancement of 
Re$\kappa_{0}^{11}(\vq,\,\omega)$ by the $d$-wave form factor 
in \eq{chid}. 
Since \eq{resonance} describes an in-gap collective mode at moderate 
$\vq$, the peak of Im$\kappa^{11}(\vq,\,\omega)$ in the 
small $q$-region is regarded as an overdamped collective mode of the 
$d$FSD.

We have investigated the RPA excitation spectrum of the $d$FSD 
within the slave-boson mean-field approximation to the 2D $t$-$J$ model. 
In a normal state, excitation spectrum shows a low energy peak, 
which connects to critical fluctuations of the $d$FSD at lower $T$. 
The competition with the $d$-wave pairing, however, blocks the 
critical fluctuations. The whole spectral weight is transferred 
to high energy and a pronounced peak appears there 
in the $d$-wave pairing state. This peak is an overdamped collective 
mode of the $d$FSD and can grow to be a resonance mode at moderate $q$.

While these results are obtained in the RPA, 
we expect that higher order corrections will not 
modify appreciably at least the results near $T=0$ (\fig{dImkqw}), 
since the boson condenses at the bottom of its band 
and the U(1) 
gauge field describing fluctuations around the mean fields is not 
relevant. 
On the other hand, the results of \fig{uImkqw} are 
obtained at finite $T$ and a $q$-linear behavior of the low energy 
peak might not be a robust property.

Correlations of the $d$FSD are ingredients of both the 
2D $t$-$J$ model\cite{yamase00} and 
Hubbard model\cite{metzner00,wegner02,frigeri02,neumayr03}.  
Their implications for high-$T_{\rm c}$ cuprates are interesting. 
Since appreciable correlations of the $d$FSD make the electron system 
sensitive to an extrinsic anisotropy between the $x$-direction and 
the $y$-direction, 
even a small anisotropy can be sufficient to lead to the $d$FSD, possibly 
a quasi-1D FS in each CuO$_{2}$ plane. 
This possibility was proposed for Nd-doped LSCO systems.\cite{yamase00}  
Fluctuations of the $d$FSD in such a quasi-1D state 
will be investigated elsewhere.\cite{yamase04c}  
In the absence of a (static) $xy$-spatial anisotropy, 
the present theory is applicable 
and we expect the excitation spectrum \fig{uImkqw}(c) in the normal state 
and \fig{dImkqw}(c) in the $d$-wave pairing state.\cite{miscspingap}  
As a direct test, however, 
conventional optical methods are not sufficient,  
since they measure a quantity with $\vq=0$ and 
will not reach a finite $\vq$-region, especially the region 
where the resonance mode is predicted. 
Indirectly, searching some phonon anomalies may be 
promising since fluctuations of the $d$FSD are expected to couple 
with a lattice degree of freedom. 
However, we do not have calculations on such  coupled 
systems at present.

Fluctuations of the $d$FSD should not be confused with 
those of spin-charge stripes.\cite{tranquada95,kivelson03} 
(i)  The $d$FSD is generated by forward scattering while formation of 
spin-charge stripes requires an interaction 
with large momentum transfer. The underlying physics is different. 
(ii) Fluctuations of the $d$FSD are relevant in systems near the 
breaking of the square lattice symmetry 
while stripe fluctuations make sense 
in systems near translational symmetry breaking.

While some hidden orders are often discussed 
in the connection with the pseudogap,\cite{timusk99} 
correlations of the $d$FSD are not related directly to the 
pseudogap in the slave-boson scheme.\cite{fukuyama98} 
However, their effects may contribute to pseudogap behaviors  
additively, since the present FS deformation has the $d$-wave symmetry, 
the same symmetry as the pseudogap.

The author is grateful to W. Metzner for  
a critical reading the manuscript and 
useful discussions, and to H. Kohno 
for helpful discussions at the early stage of this work. 
He also thanks L. Dell'Anna  
and R. Zeyher for stimulating discussions. 
A part of this work was supported by 
a Special Postdoctoral Researchers Program from RIKEN.     

\bibliography{main.bib}

\begin{thebibliography}{30}
\bibitem[{mis({\natexlab{a}})}]{miscdFSD}
\bibinfo{note}{A stability condition of the Fermi surface was discussed by I.
  J. Pomeranchuk, Sov.\ Phys.\ JETP {\bf 8}, 361 (1958). In this sense, the
  $d$FSD is often referred to as Pomeranchuk instability.}
\end{thebibliography}

\end{document}